# A Scalable and Interoperable Platform for Transforming Building Information with Brick Ontology

**Rozita Teymourzadeh, PhD, CEng.**
*Azbil North America Research and Development, Inc.*
*Senior IEEE, IET*

**Yuya Nakazawa,**
*Azbil Corporation,*
*AI Solution Department*

**ABSTRACT**

*In the digital twin and building information era, many building automation companies searched for scalable methods to extract and analyze different building data, including Internet of Things (IoT) sensors, actuators, layout sections, zones, etc. The necessity for engineers to continuously manage the entire process for each new building creates scalability challenges. Furthermore, because construction information is sensitive, transferring data on vendor platforms via the cloud creates problems. This paper introduces a platform designed to address some of the common challenges in building automation. This is a smart platform designed for the transformation of building information into Brick ontology (Brick 2020) and graph formats. This technology makes it easy to retrieve historical data and converts the building point list into a Brick schema model for use in digital twin applications. The overarching goal of the proposed platform development is semi-automate the process while offering adaptability to various building configurations. This platform uses Brick schema and graph data structure techniques to minimize complexity, offering a semi-automated approach through its use of a tree-based graph structure. Moreover, the integration of Brick ontology creates a common language for interoperability and improves building information management. The seamless and offline integration of historical data within the developed platform minimizes data security risks when handling building information.*

**INTRODUCTION**

Building automation companies have actively pursued scalable solutions to efficiently extract and analyze a wide spectrum of building data in the dynamic world of digital twins and building information. Many previous attempts to address challenges in the realm of building information were hindered by the lack of success, mainly attributed to concerns about scalability (Rosen et al. 2015). The overarching issue required a common language for sensors in buildings to eliminate redundancy. The industries of Building Management Systems (BMS) and Building Information Modeling (BIM) (Wang and Xie 2002) acknowledged the lack of sufficient information in the BIM and Industry Foundation Classes (IFC) files (Vittori 2023). IFC files included building architectural and spatial data; however, they did not include sensor and actuator data. According to Balaji (Balaji et al. 2016), the National Institute of Standards and Technology (NIST) highlighted the critical problem of a missing common data representation, which obstructs interoperability between buildings and hampers the scalability of applications. Developers were forced to manually map heterogeneous data from each building to a common format, leading to inefficiencies. In response to these challenges, the Brick community, in 2016 (Balaji et al. 2016), introduced semantic web

**Rozita Teymourzadeh** is assistant professor and Director at Azbil North America Research and Development, Inc. Santa Clara, California, USA. **Yuya Nakazawa** is an AI researcher at Azbil Corporation, Fujisawa, Japan.

technology as a solution. This initiative aimed to standardize semantic descriptions of physical, logical, and virtual assets within buildings (Brick 2020), including their interrelationships. Despite this advancement, current data integration processes still involve substantial manual interventions, indicating a need for further optimization to achieve seamless interoperability and scalability in the representation of building information from the source of creation. We thoroughly scrutinized existing ontologies such as Building Topology Ontology (BOT) (Rasmussen et al. 2021), Semantic Sensor Network/Sensor, Observation, Sample, and Actuator (SSN/SOSA) (Haller et al. 2017), Smart Applications REFerence (SAREF) (Garcia et al. 2023), Real Estate Core (REC) (Hammar et al. 2019), Project Haystack (Quinn and McArthur 2022), Brick ontology (Brick 2020) as well as proposed draft version of ASHRAE Standard 223P (ASHRAE 2018). This detailed research serves as a foundational step in our pursuit of improving and optimizing existing ontological substructures. While we recognize the inherent benefits and downsides of various ontologies in the context of building automation, we chose Brick ontology as the foundation of our development efforts. This decision was made after careful consideration of the unique characteristics, adaptability, and alignment with the specific requirements that the Brick ontology provides. Brick schema (Balaji et al. 2016) offers several benefits that justify our decision to put this ontology on the shortlist for our continuous research work. The main advantages are listed below.

- Standardized Semantic Descriptions and Relationships.
- The schema is built on contemporary Semantic Web technologies, using the Resource Description Framework (RDF) (McBride 2004) and the Web Ontology Language (OWL) (Grigoris and Harmelen 2009). This modern foundation contributes to the schema's robustness and adaptability.
- The Brick schema facilitates the scalability and portability of controls, analytics, and other applications within the industrial context, ensuring flexibility in deployment.
- The schema adeptly captures both explicit and implicit relationships necessary for a spectrum of applications, enhancing its versatility.
- Brick encompasses descriptions of pertinent concepts essential for diverse applications, aligning with the varied needs of users.
- The schema caters to the requirements of both domain experts and application developers, fostering collaboration and meeting diverse perspectives.
- Brick enables an automatic validation process to ensure the correct usage of its concepts, enhancing data integrity and reliability.
- The schema supports the use of templates for standard equipment, streamlining the integration process, and ensuring consistency in representation.

These benefits combine to make Brick a compelling solution, in line with our goals of efficient and standardized ontology utilization in our industrial application. By using Brick as our fundamental framework, we aim to rely on its well-defined structure and comprehensiveness, providing us with a solid scaffolding to on which to build. This choice is the product of a rigorous and intentional process that ensures our approach to building automation is not just technically solid but also optimally aligned with the domain's complexities and nuances. Following the introduction of Brick in 2016 (Balaji et al. 2016), numerous research initiatives have emerged aimed at enhancing the compatibility of Brick with other ontologies and bridge the gap between manual and automated processes. An illustrative example is the work by Fierro (Fierro et al. 2020). Their study provides a qualitative analysis of Project Haystack, focusing on a tagging system for building metadata. Through this analysis, they identified a set of inherent definable and consistency issues within the tagging model. In response, the research aimed to present a solution by proposing a replacement of Brick with clear formal semantics. Exploring Brick metadata persisted, and Fierro (Fierro et al. 2019) took the initiative to harmonize additional ontologies such as Haystack (Haystack 2018) with the existing Brick ontology. Haystack, recognized for its tagging system, became the focal point of a qualitative analysis presented by the author. This analysis delved into the intricacies of Haystack project (Haystack 2018) as applied to building metadata. In particular, the authors introduced Brick ontology, characterized by lucid formal semantics. This innovative addition facilitates inferring a valid Brick model from an initially informal Haystack model. The research extends beyond mere exploration, actively contributing to the refinement and integration of ontologies for enhanced efficacy in building metadata representation. Garrido (Garrido-Hidalgo 2022) brought forth the idea of interlinking the Brick schema with a building domain ontology. Their proposal involved the development of an interactive ontology matcher that has weak supervision and active learning. Although the approach appears to be logical, a notable gap lies in the absence of a comprehensive test report that provides in-depth accuracy of the platform. In 2022, Fierro (Fierro et al. 2022) put a principle on the construction of semantic metadata models, introducing the concept of semantic sufficiency. According to this principle, a model is considered "finished" when it incorporates the metadata essential to support a specific set of applications. The methodology involves soliciting feedback from application metadata requirements and employing a templating system to generate common metadata model components. Although this approach has the potential to enhance metadata precision, it involves a trade-off by introducing customized manual work for specific requirements. The emphasis is on striking a balance between increased precision and the inclusion of tailored elements, acknowledging the nuanced demands of diverse applications. During a short period of time, multiple research efforts have been carried out to enhance the utilization of Brick Metadata (Lee et al. 2023). Developers have actively desired to bridge the gap between automated and manual processes, introducing additional tools and utilities. Consequently, this paper presents a scalable solution designed with the Brick metadata ontology as its foundational structure and the backbone of the proposed platform. The primary objective of the developed platform is to effectively retrieve data from building sensors and actuators, converting this information into semantic point label information compatible with Brick ontology and scalable for existing and new buildings. This transformation facilitates the classification and standardization of building data using the Brick ontology, allowing the easy development of applications such as energy monitoring and data visualization. The subsequent sections delve into a more detailed exploration of the implementation process, shedding light on the various facets of the platform.

**PROPOSED PLATFORM INTRODUCTION**

The inception of the proposed platform was driven by the need to fully capture building data, including sensor status, IoT sensors, actuators, and various digital and analog data generated by building automation systems. Typically, acquiring access to the building gateway or the BACnet system (Bushby et al. 2002) involves navigating through an administrative process and seeking permissions from multiple departments. This procedure is not only time-consuming, but, at times, permissions are withheld due to concerns surrounding data security. To address this challenge, we opted for an alternative approach. Instead of directly connecting to the live system, we decided to retrieve and process historical data spanning a single year. This data is extracted, and the platform is parsed through the stored data offline and precisely analyzed the historical data. This method not only avoids the complexities associated with permission but also ensures a secure and efficient means of conducting our analysis in terms of data analysis and building data normalization. Our focus has been on accumulating a substantial dataset over an entire month and year from a building located in Azbil Fujisawa Technology Center (FTC), Japan (Azbil Corporation 2024) hereafter referred to as the FTC building. This building comprises six floors, with a total area of 2,320 square meters (24,972 square feet) and a total floor space of 10,721 square meters (115,399 square feet). For illustrative purposes, the following figure presents an example of the data, classified as point labels and time series data, extracted from the FTC building. This project demonstrates our commitment to creating a robust platform that is suited to the specific needs of real-world building scenarios. It is noted that the point label data presented in this document have been modified from their original values to ensure data privacy.

| Point label ID | code | 12.34.567.890 | 12.34.567.891 | 12.34.567.892 | 12.34.567.893 |
|---|---|---|---|---|---|
| Point label | name | VAV-1 給気温度設定値 | VAV-1 給気温度計測値 | VAV-2 給気温度設定値 | VAV-2 給気温度計測値 |
| Other data associated with point label | unit | °C/°F | °C/°F | °C/°F | °C/°F |
| Time series value | 4/1/2023 | 25.0/77.0 | 24.8/76.64 | 25.0/77.0 | 25.2/77.36 |

**Figure 1**   An example of point list and time series data

Using this technique to gather historical building data not only improves security but also acts as a preventative precaution against possible cyber threats. By avoiding direct access to real-time data, we reduce the likelihood of harmful hits on the real-time system. This precautionary measure ensures a strong and secure framework that protects the integrity and confidentiality of building data throughout our analysis operations. After preparing the data for processing using Brick ontology (Brick 2020), the following section will focus on the design architecture that accommodates the platform.

## PLATFORM DESIGN ARCHITECTURE

THE proposed platform is a semi-automated system that takes building information in raw, non-standard formats and transforms it into a standardized version that is compatible with the Brick schema. The principal objective is to correct the data from the source. Despite these advancements, our development journey encountered several challenges:

1. Limited access to building information due to the absence of a connection to the BACnet system.
2. The variable nature of building information presents a diversity of data structures and formats.
3. Language barriers, particularly in dealing with point label data generated in Japanese language.

In response to these complications, this platform evolved as a smart solution in the field of building automation. It deliberately tackles these challenges by effortlessly converting building information into Brick ontology and graph formats. This modification improves scalability in building automation and allows for a faster deployment procedure across several buildings, reducing the need for considerable human intervention by engineers. A key aspect of this platform is its ability to match point lists with the Brick schema using a classifier model trained on Artificial Intelligence (AI) within its backbone, organized in a dictionary format. This platform has three main duties that contribute to its powerful functionality:

- Data Preparation: This involves translation and normalization processes that ensure that the data are formatted and aligned for further analysis.
- Point Label Mapping: Using AI and a purpose-designed algorithm, the platform undertakes the crucial task of mapping point labels between the current building information and the corresponding Brick class and sub-classes.
- Graph Model Generation: generating a comprehensive graph model based on Brick metadata, providing a structured representation of building information.
- Graph Validation: The platform employs validation mechanisms to ensure the accuracy and integrity of the generated graph model.

This platform offers the following underlying architecture to create graph (McBride 2004) from the point labels. Refer to Figure **2** for a visual representation of the top-level architecture of the platform. The subsequent section describes the building data transformation into the Brick schema model.

**Preparation and Dependencies:**

During this essential phase of our project, we systematically build up dependencies, integrate necessary libraries, and populate the system with necessary data. An important part of this phase is data translation, which converts sensor data and semantic representations of sensors and actuators from Japanese to English language. This translation is required for an accurate matching to Brick classes. To help with this procedure, we used and update a translation service with a built-in Japanese dictionary. Figure 3 illustrates an example of the dictionary module used in this project. Furthermore, a key component is the integration of the Building Metadata Ontology Interoperability Framework (BuildingMOTIF) (The National Renewable Energy Laboratory (NREL) 2023), a versatile toolkit that facilitates the creation, storage, visualization, and validation of building metadata. Delivered as an SDK with APIs, BuildingMOTIF abstracts complexities related to RDF (McBride 2004) graphs, database management, Shapes Constraint Language (SHACL) (W3C 2016) validation, and interoperability across diverse metadata schema and ontologies. This project utilizes the Brick ontology's "Nightly_1.3" version and an AI-trained data model, both loaded during the preparation phase. It is worth noting that due to security considerations and the unavailability of building layout and mechanical specifications, the proposed platform uses the point label information to construct the semi-layout in JavaScript Object Notation (JSON) format. This semi-layout, although not exceptionally precise, serves its purpose well, enabling the detection of the required number of floors and rooms for our project. For a visual representation, refer to Figure 4 that illustrate an example of the layout generated for the FTC building.

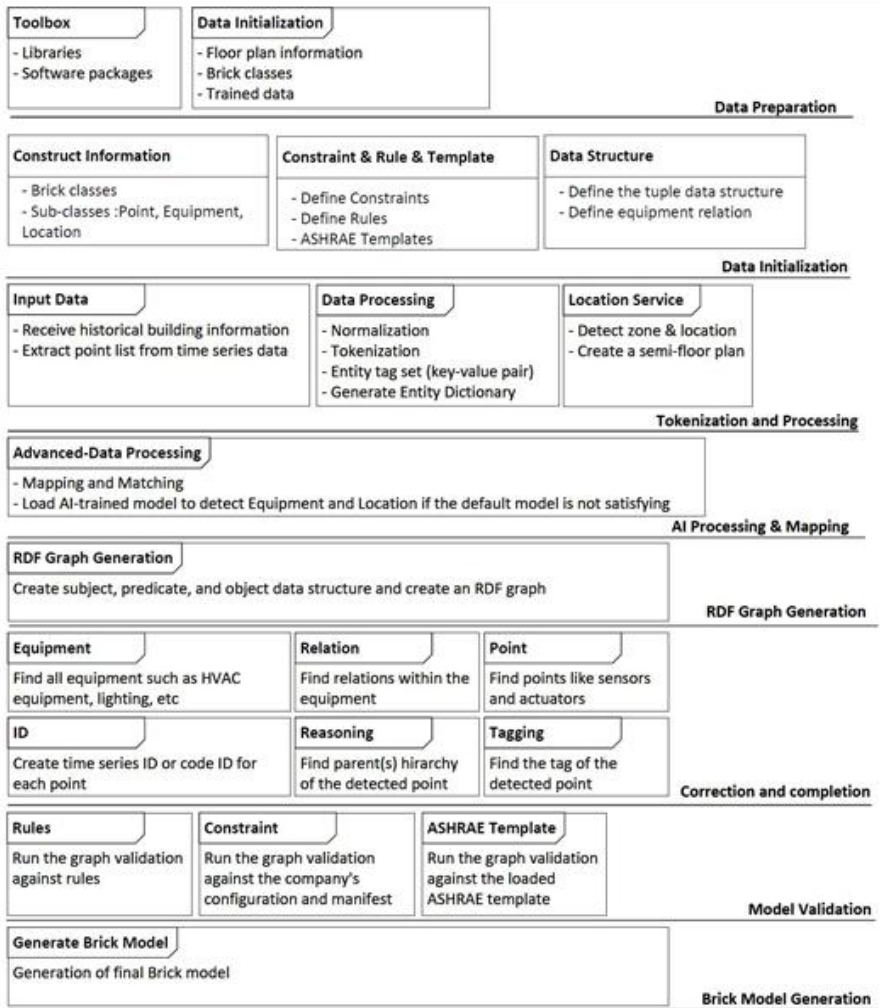

**Figure 2    Proposed top-level platform architecture**

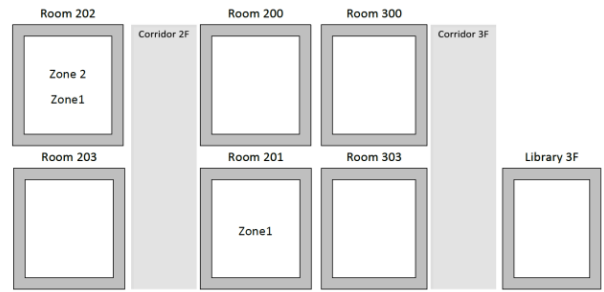

Figure 3 Example of dictionary service

Figure 4 Building semi-layout generated by point label

**Initialization:**

During this step, the extraction, sorting, and evaluation of Brick classes take precedence, serving as the foundation for further mapping and AI job processing. Here, the data structure rules for transforming point labels into graphs are thoroughly described. The fundamental data structure of RDF graph revolves around the interaction of subjects, objects, and predicates (SOP) that are linked to shape the RDF graph (McBride 2004) tuple model. This hierarchical arrangement contributes significantly to the comprehensive representation of the building. The subsequent model serves as an illustrative example, focusing on an element of an FTC building (Azbil Corporation 2024), specifically within the Heating, Ventilation and Air Conditioning (HVAC) system (Trcka et al. 2010). Figure 5 shows some examples of tuple data structure:

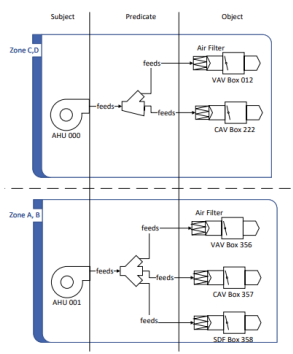
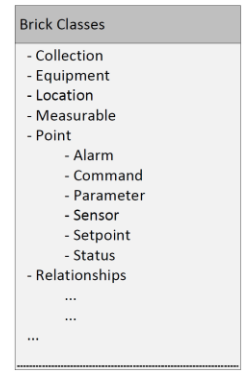

**Figure 5 Tuple data structure of targeted FTC building**   **Figure 6 Example of the Brick class (Brick 2020)**

This information is compiled by extracting point labels, normalizing the data, and generating a tuple data structure for each equipment connected in the building.

**Tokenization and Data Processing:**

In this section, the point label undergoes a thorough data-cleaning process employing several algorithms to detect and extract pertinent information. The extraction includes key details, including: (1) Information about zones, floors, and areas. (2) First iteration of Brick class detection. (3) Device information. (4) Relations and connections.

The organized data are then structured into key-value pairs for Brick classes and lists for device information. This structured format serves as input for the upcoming phases, specifically AI processing and detection, resulting in a comprehensive approach to data utilization in the next stage of our project. This outcome prompts us to create a semi-layout and the starting point for the introduction of equipment connection points.

**Mapping and AI Processing:**

The mapping and AI processing stages play an important role in this chain process. Mapping involves identifying similar Brick classes for each point label generated by the building data. This proves to be a challenging task due to differences in semantics, point labels, and the digital representation of data. In a 2020 study (Fierro et al. 2020), Fierro proposed a unified authoritative Brick Metadata model for buildings, intended for continuous maintenance throughout the life cycle. However, the lack of extensive building data for research limited the testing of these algorithms. In our research, the Japanese representation of building metadata serves as a valuable source of point label data for our matching purposes. In the Brick ontology, the ideal outcome is a match in the form of a key-value pair for Brick classes. For a visual reference, Figure 6 above illustrates examples of Brick classes, sub-classes, and their relations. This phase lays the support for accurate mapping and processing, overcoming the challenges associated with semantic variations in the data. Point labels are created by putting the building data through a rigorous process of translation, normalization, and cleaning to align them with Brick classes.

Starting with a trained predetermined model that is organized into key-value pairs, the mapping algorithm begins. Later, an AI algorithm updates this predefined model with a model taught. To efficiently train the data and then map it with the intended building point list, we have investigated several AI techniques and frameworks to train the data such as Random Forest (Briman 2001), Active learning text-based Scrabble framework (Koh et al. 2017), and Large Language Models (LLM) framework like OpenAI (Brown et al. 2020). To validate our generated graph model, we utilized various building datasets for comparison and benchmarking. These include Ecobee dataset (Ecobee 2024), the High-Fidelity Building Emulator (High-Fidelity Building Energy Data Collection 2024), the Honda Smart Home (Honda Smart Home Data Collection 2024), the Lawrence Berkeley National Laboratory Building (Building Data Genome Project 2024), among others. The condensed model is kept on the platform in dictionary format. With the integration of Brick schema, this research methodology helps the algorithm in interpreting, mapping, and classifying the intended building information. Later, we used Jaccard index (Skiena 2008) to measure the similarity between two sets of normalized trained point label and Brick point and location classes. The Jaccard index is calculated using the following equation:

$$J(Set_A, Set_B) = \frac{|Set_A \cap Set_B|}{|Set_A \cup Set_B|} \quad (1)$$

The Jaccard index is used to estimate the best match similarity between the point and the Brick classes. Notably, the raw point data has already been normalized using an LLM-trained module. The following table shows an example of mapping results of the matching algorithm.

Table 1.  Brick class and point label match example

| Brick Class | Point label | Algorithm Match |
|---|---|---|
| Temperature_Sensor | SDF_65_Zone_Average_Temp | Average_Zone_Air_Temperature_Sensor |
| Occupancy_Count_Sensor | SDF1_People number | Occupancy_Count_Sensor |
| Humidity_Sensor | AHU_67_Indoor_Humi | Humidity_Sensor |
| Illuminance_Sensor | SDF_3_WP_Sensor_Illuminance | Illuminance_Sensor |
| Auditorium | VAV_6_F_Audience_Area | Auditorium |
| Not applicable | Reserve__AV | No Match |

We evaluated about 7800 labels and about 7400-point labels were matched with Brick classes. The remaining 400 points were either "Reserve" points or lacked sufficient information to match with a Brick class. Enhancements to our AI matching algorithm are ongoing but beyond the scope of this paper.

**RDF Graph Generation:**

This stage is important within the algorithm. As discussed earlier, the data have been normalized, mapped, and classified. In this phase, we established a tuple data structure for the converted building data, now referred to as point label. Although the point label contains information about the Brick class, it does not define a relationship within the building as a reference to other elements. To bridge this gap, each point label needs to be converted into a data structure of object, predicate, and subject, as illustrated here:

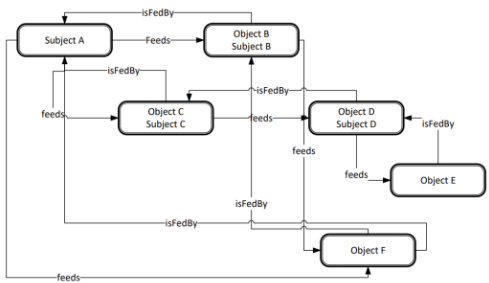

Figure 7 Point label in Brick RDF graph format                Figure 8 Thermostat template with CO2

Iterating through each point label that is accessible, assigning a matching Brick class, and constructing the relationship to turn the point label into the Brick RDF graph are all part of the loop process. The RDF graph has a tree-like shape and is widely considered the best option for knowledge graphs due to its native web syntax, which makes data sharing and exchange easier (Antanas 2022). Additionally, because of its formal semantics, meaning and structure can be easily aligned across many sources, resulting in unified perspectives and clear interpretation. The tree-like data structure of the RDF makes it suitable for web queries and allows the achievement of a linear time and space query. A graph-format data structure is attractive for storing and accessing point labels and constructing information in databases because of its advantages, especially when used with RDF. The RDF is hierarchical, making it compatible with the SPARQL (Furche et al. 2010; DuCharme 2013) and GraphQL (GraphQL 2021) query languages and making fast data retrieval possible. All in all, these benefits put RDF in a strong position for constructing information within a database architecture, as well as for deploying and querying point labels.

**Correction and Completion:**

At this point, the algorithm adds relations to a previously discovered equipment and integrates it to the main graph, the algorithm searches for air handling unit (AHU) terms (we will talk about AHU terms in the upcoming chapter) to make sure no unit is overlooked. Depending on whether the mismatched point label is produced, it will be either removed from the graph or matched with the closest matching factor and included in the main graph in this review part. Here, reasoning and inference (Brick 2020) are implemented. Inference and reasoning are processes to imply information from the Brick class and expand it to the main graph. Based on the application requirement, detailed information is added to the graph as a separate branch. This information includes super-class, inverse relation, and tagging information.

**Model Validation:**

Now the graph is generated using the point label and reviewed, and the missing element is added or removed from the graph according to their matching algorithm through the reasoning process. In this section, we check the generated graph against constraints and guidelines. The idea of graph validation is inspired by BuildingMOTIF development (NREL 2023). During the validation process, we specify the equipment to be validated using predefined templates. These templates are designed to ensure the accuracy of the generated Brick model, with the final report produced using the BuildingMOTIF library (NREL 2023). Each template delineates the required components and their dependencies, such as setpoints, temperature points, and CO2 points. An example of such a template is provided in **Figure 8** (NREL 2023).

**Brick Model Generation:**

This stage finalizes the architecture by generating the Brick model. At this stage, the data taken from the FTC building point list is organized into a tuple data format and then converted into a Brick model. Building data, both dynamic and static, are transformed into a machine-readable, semantic format by this process. A format like this makes it easier to create queries for sensors and actuators, which makes it possible to retrieve data efficiently for monitoring applications.

**DETAILED ARCHITECTURE DESIGN**

The project architecture starts with the translation and parsing utility for the point label. To align the raw data point label with the triple SOP data structure, the data goes through several steps; after processing the data, it is correlated with the relevant Brick class model to produce an RDF Brick model. The architecture's block diagram is displayed in Figure 9. The mapping stage in point processing involves intensive AI processing and data normalization to align the point list with the appropriate Brick class. Subsequently, it is necessary to identify HVAC-related terms within the dataset, as these terms are important to define relationships within the list of points. HVAC term detection is carried out through input from an interactive user interface specifically designed for this purpose. For example, the following point refers to HVAC components: (1) AC: Air handling unit (AHU) (2) SDF: Smart-control-damper on the Diffuser (3) VAV: Variable air volume

(4) CAV: Constant air volume. The implemented mapping algorithm analyzes the HVAC elements within the point list and determines the relationships between them. Figure **10** illustrates the defined relations and their reverse relations for AHU terms, while Figure 11 displays the interactive graphical user interface designed to capture operator input.

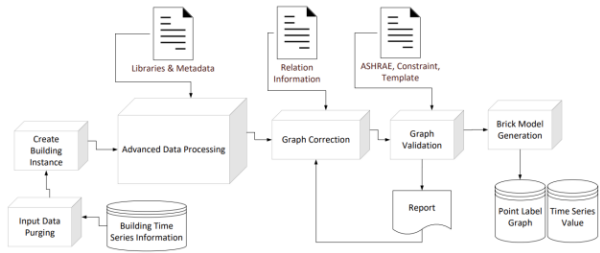
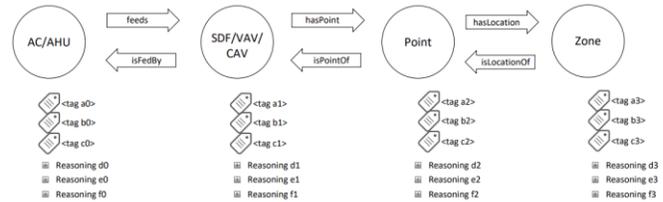

Figure 9 The developed platform architecture       Figure 10 FTC building AHU terms relation

To enhance scalability and streamline debugging, the algorithm divides each equipment unit and its associated relationships into separate modules. Later, these modules are combined to generate a comprehensive model representing the building. For example, relationships like AC to CAV and AC to SDF are handled in individual subsections (**Figure 11**, **Figure 12**):

- AC2CAV: AC and CAV units and relevant Brick relations.
- AC2SDF: AC and SDF units and relevant Brick relations.
- AC2VAV: AC and VAV units and relevant Brick relations.
- CAV2SDF: CAV and SDF units and relevant Brick relations.
- Point connection: Point connection to the zone or other unit equipment.
- Tagging: Create relevant tags for a specific point.
- Reasoning: Retrieve the parent class of the specific point.

Figure 12 illustrates this structure in a scalable module, which reduces processing time when certain modules are not required. Each module can be enabled or disabled based on the application's needs.

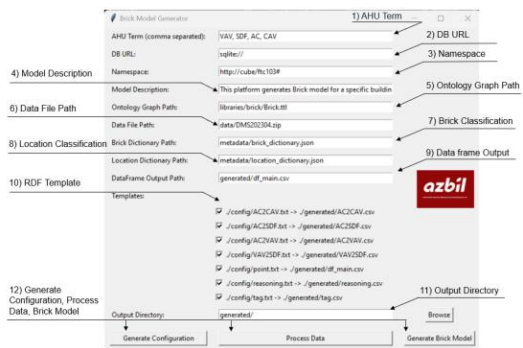
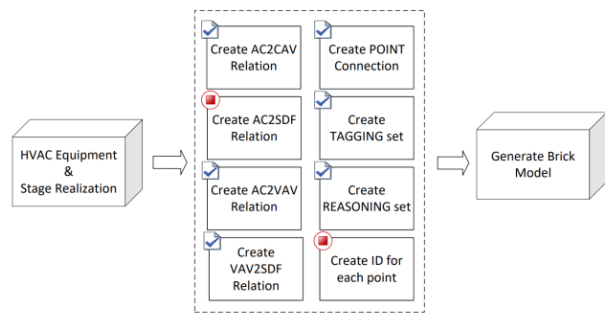

Figure 11 The front-end application for data processing       Figure 12 Detailed architecture to generate Brick graph

Here the processing and data normalization is concluded, and the Brick graph is generated and stored in the database for data access processing. In the next section, we will discuss the result in detail.

## RESULT AND DISCUSSION

Following the processing and mapping of each point to the relevant Brick class, and defining the relationships between equipment units, the Brick module graph is generated and combined to represent the semantic version of the building data. It also includes sensors and actuators information. The following relationships were implemented in this module (**Figure 13**) while output of the developed platform which represents the Brick module of FTC building is shown here (**Figure 14**):

**Implemented Relations**
- feeds
- isFedBy
- isPointOf
- hasPoint
- hasTag
- hasTimeseriesID
- ...

```
Brick model representation of a locker in the library
@prefix brick: <https://brickschema.org/schema/Brick#> .
@prefix ftc103: <http://cube.com/ftc103#> .
@prefix ref: <https://brickschema.org/schema/Brick/ref#> .

ftc103:10F_536_Locker_Room_in_library_On_Off_Status a brick:On_Off_Status,
        brick:On_Status,
        brick:Point,
        brick:Status ;
    brick:hasTag <tag:Off>,
        <tag:On>,
        <tag:Point>,
        <tag:Status> ;
    ref:hasTimessriesId "2340267377769557870291170063628324592333" .
```

Figure 13 Brick relation implemented in the platform     Figure 14 Generated Brick model for FTC building

In this graph, three prefixes were extracted: Brick, FTC building, and reference for the time series ID. The first point represents the tuple structure for the status of locker 536, located in the library. The status value of this locker is stored in an SQL database, which can be retrieved upon receiving a query. This point includes reasoning attributes like "On_status", "Point", and "status", and tags such as "On", "Point", and "Status". Another example, shown in Figure 15, illustrates a tuple object referring to AHU unit No. 977 (AC 977), which supplies specific CAV and SDF units as listed below.

```
AHU to CAV/SDF relations
ftc103:AC_977 a brick:AHU ;
    brick:feeds ftc103:CAV_635,
        ftc103:SDF_223,
        ftc103:SDF_445,
        ftc103:SDF_154,
        ftc103:SDF_137,
        ftc103:SDF_488,
        ftc103:SDF_987,
        ftc103:SDF_234,
        ftc103:SDF_128,
        ftc103:SDF_444 .
```

```
AHU and POINT relations
ftc103:AC_600 brick:hasPoint ftc103:3F_VAV_331_System__AC_303__Status_On_Off,
    ftc103:10F_VAV_811_System__AC_600__TM_Status,
    ftc103:10F_VAV_811_System__AC_600__TM_On_Off,
    ftc103:10F_VAV_812_System__AC_600__TM_Status,
    ftc103:10F_VAV_812_System__AC_600__TM_On_Off,
    ftc103:10F_VAV_811_System__AC_600__TM_Status,
    ftc103:10F_VAV_811_System__AC_600__DAMP_On_Off,
    ftc103:10F_VAV_811_System__AC_600__DAMP_Status,
    ftc103:10F_VAV_811_System__AC_600__VAV_Inlet_Temp,
    ftc103:AC_600_Max_Min_Value_Calculated_Chilled_Water_Interlock,
    ftc103:AC_600_Max_Min_Value_Calculated_Hot_Water_Interlock,
    ftc103:AC_600__Interior_etc_SA_Total_On_Off_Status,
    ftc103:AC_600__Interior_etc_SA_Total_Static_Pressure .
```

Figure 15 AHU unit has "feeds" relation to the CAV and SDF units     Figure 16 **AHU** unit and **point** relation

In this example Figure 16, the object created from the points represents AHU unit AC 600, which feeds the zone equipped with the temperature sensor, damper, and the points listed here. Finally, the generated Brick model provides a detailed understanding of the sensors deployed in the building, such as a WP (Workplace) sensor of the wireless cell-type HVAC system installed in the FTC building as shown in the following figure. Figure 17 shows the WP (Workplace) sensor detected by the Brick model and is part of the building automation system.

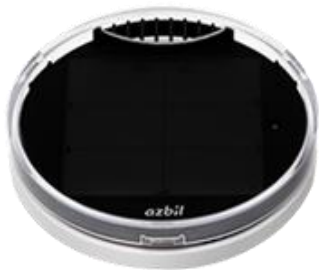

```
WP (Workplace) sensor representation in generated model
ftc103:SDF__102_7 brick:hasPoint ftc103:SDF_102_7_SP_Value,
    ftc103:SDF_895_WP_Sensor_Device_ID,
    ftc103:SDF_895_WP_Sensor_Illuminance_Data,
    ftc103:SDF_895_WP_Sensor_Primary_Battery_Level .
```

Figure 17   Generated Brick model explains WP (Workplace) sensor

These output examples from the generated Brick module on the platform demonstrate the relationships between equipment and units across different building sectors, highlighting the model representation's scalability. The platform was tested in several buildings and only minor adjustments were needed. The generated graph and time series data are stored in a database and can be used for data access, energy monitoring, and data visualization applications.

## CONCLUSION

This article discussed the necessity of decreasing human labor and the scalability issues with contemporary building automation. With the help of the developed platform and Brick ontology, we suggested a scalable and seamless solution. Using historical building data, this platform creates a Brick model graph that captures and maintains the links established by the Brick ontology. Examples of this type of information include sensors, actuators, and the time-series data that goes along with them. We presented an overview of the developed architecture platform, and several examples of the generated Brick output models were illustrated and discussed in detail. The developed platform has been tested with multiple buildings, and the Brick graph data model was generated successfully. The proposed platform produced realistic graphs that reflected the operational data and the semi-layout of each building. The scalability of this solution makes it possible to automatically generate machine-readable semantic building models, greatly minimizing the manual labor needed to produce these models from scratch. Additionally, the platform creates a uniform framework for expressing building information by integrating Brick ontology, making interoperability and effective data management between diverse building automation systems possible.

## ACKNOWLEDGMENTS

This project was developed by Azbil North America Research and Development Inc., with support from the Azbil AI Solution department. I sincerely thank everyone involved, especially Dr. Gabe Fierro, founder and lead maintainer of Brick Schema; Chosei Kaseda and Jeremy Tole from Azbil North America Research and Development, respectively for their invaluable guidance and recommendations throughout the project.